\documentclass[conference]{IEEEtran}
\IEEEoverridecommandlockouts

\usepackage[utf8]{inputenc}
\usepackage{amsmath}
\usepackage{color}
\usepackage{graphicx}
\usepackage{url}

\usepackage[siunitx]{circuitikz}
\usepackage{subcaption}

\usepackage[acronym]{glossaries}
\setacronymstyle{long-short}
\newacronym{st}{S/T}{Schmitt-Trigger}

\usepackage{textpos}
\usepackage{hyperref}

\begin{document}

\title{The Metastable Behavior of a Schmitt-Trigger}
\author{
\IEEEauthorblockN{ Andreas Steininger
\begin{minipage}[c]{1em}
\href{https://orcid.org/0000-0002-3847-1647}{{\includegraphics[width=1em]{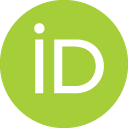}}}
\end{minipage}\,
and\: J\"urgen Maier
\begin{minipage}[c]{1em}
\href{https://orcid.org/0000-0002-0965-5746}{{\includegraphics[width=1em]{orcID.png}}}
\end{minipage}\,
and\: Robert Najvirt}
\IEEEauthorblockA{Vienna University of Technology, 1040 Vienna, Austria\\
\{steininger, jmaier, rnajvirt\}@ecs.tuwien.ac.at}
\thanks{This research was partially supported by the SIC project (grant
  \mbox{P26436-N30}) of the Austrian Science Fund (FWF).}
}

\maketitle

\begin{textblock*}{\textwidth}(0mm,195mm)%
  \footnotesize%
  \textcopyright\ 2016 IEEE.  Personal use of this material is permitted.  Permission from IEEE
  must be obtained for all other uses, in any current or future media, including
  reprinting/republishing this material for advertising or promotional purposes,
  creating new collective works, for resale or redistribution to servers or lists,
  or reuse of any copyrighted component of this work in other works.
\end{textblock*}%
\vspace*{-1em}

\begin{abstract}
  Schmitt-Trigger circuits are the method of choice for converting general
  signal shapes into clean, well-behaved digital ones.  In this context these
  circuits are often used for metastability handling, as well.  However, like
  any other positive feedback circuit, a Schmitt-Trigger can become metastable
  itself.  Therefore, its own metastable behavior must be well understood; in
  particular the conditions that may cause its metastability.

  In this paper we will build on existing results from Marino to show that (a) a
  monotonic input signal can cause late transitions but never leads to a
  non-digital voltage at the Schmitt-Trigger output, and (b) a non-monotonic
  input can pin the Schmitt-Trigger output to a constant voltage at any desired
  (also non-digital) level for an arbitrary duration.  In fact, the output can
  even be driven to any waveform within the dynamic limits of the system.  We
  will base our analysis on a mathematical model of a Schmitt-Trigger's dynamic
  behavior and perform SPICE simulations to support our theory and confirm its
  validity for modern CMOS implementations.  Furthermore, we will discuss
  several use cases of a Schmitt-Trigger in the light of our results.
\end{abstract}

\section{Introduction}
\label{sec:introduction}

It is a fundamental task in digital computation to discriminate the analog
voltage levels carried by the signal rails in the physical implementation in two
logical classes, namely those representing a logic HI and those representing a
LO.  That can normally be managed by the conventional input stages of logic
gates. However, when there is a need for handling less ``clean'' signals with
intermediate voltage levels, slow transitions, or large noise, special
provisions are required. This may happen at interfaces or when external
disturbances come into play, or in case of metastability of an internal bistable
element which can also be caused by clean but badly timed signals. The standard
solution for this is the use of a \gls{st} circuit. Unlike a plain discriminator
circuit that uses just a single constant reference voltage $V_T$ for separating
into HI (above $V_T$) and LO (below $V_T$) digital values, the \gls{st} exhibits
a hysteresis at its input by switching the reference voltage between $V_H$ and
$V_L$ (with $V_H>V_L$) in dependence of its current output state, with $V_H$
being applied when the output is LO and $V_L$ for a HI output\footnote{For the
  conceptual part of our analysis we consider a non-inverting \gls{st}, while
  later, in context with the practical design we will study its inverting
  version that is easier to implement.}.  This facilitates stability against
noisy input voltages in the proximity of the threshold that typically cause the
discriminator to oscillate.

Clearly, the original intention of the \gls{st}, namely to discriminate a
continuous input voltage space into two sub-spaces, does not imply a stateful
behavior. However, the hysteresis behavior desired for noise immunity does. This
caused some uncertainty about whether a \gls{st} can become metastable. Thanks
to the results of researchers like Marino \cite{Marino77} and Chaney
\cite{Chaney79} it is today clear that a \gls{st}, like any other circuit
relying on positive feedback, cannot be protected from metastability and will
therefore exhibit irregular behavior for some input voltage traces.
Still \gls{st}s are sometimes proposed for filtering metastable outputs of
bistable elements \cite{PS13}, or for uniquely classifying the logic level of a
node that is intentionally left floating for some time in order to leverage the
parasitic capacitance as a dynamic storage element \cite{Nystroem,
  Greenstreet}. So one may ask whether such approaches can actually work. In
other cases (e.g.~\cite{web}), it is hoped that for input voltages with
restricted dynamics a \gls{st} will never experience metastability. Again
something to check for in more detail.

In this paper we extend existing results -- mainly those from Marino
\cite{Marino77} -- to answer some of these questions that frequently plague
designers in practice. To this end we will, after giving a background in
Section~\ref{sec:background}, characterize the metastable behavior of the
\gls{st} in detail and compare it to that of a typical bistable element (e.g.\
latch) in Section~\ref{sec:analysis}. Since metastability is usually a very rare
phenomenon that eludes an experimental evaluation, our aim is to give
theoretically well founded answers and particularly identify those conditions
under which metastability of the \gls{st} can be ruled out for sure. Here we
will investigate different scenarios like monotonic and slowly changing inputs.
Next, in Section~\ref{sec:evaluation} we will validate our theoretical results
by selected SPICE simulations.  In Section~\ref{sec:usecases} we will
investigate concrete use cases of a \gls{st} in the light of our findings.
Finally, in Section~\ref{sec:conclusion} we will conclude our paper.

\section{Background}
\label{sec:background}

\subsection{Metastability}
\label{sec:metastability}

Metastability is the phenomenon when a bistable element persists in an unstable
equilibrium, the metastable state, for a prolonged time. The existence of a
metastable state is a fundamental property of every bi- or multistable system --
between every two stable equilibria there necessarily is an unstable
equilibrium. The difference lies in the behavior when the equilibrium state is
slightly disturbed: The system would return to a stable state, however, upon the
slightest disturbance from a metastable state, the latter is left in favor of
either of the stable states.

It is well understood \cite{Mar81} that every bistable element can be brought to
a metastable state in which it may rest for an unbounded time.  The
manifestation of the metastable state can be oscillation or ``creeping''
\cite{Kacp88}.  In the creeping case, which is more relevant here, we know that
the classical bistable storage elements (latch, Muller C-element) drive their
output at first to a specific ``metastable'' voltage level $V_{meta}$, where it
stays for an unbounded amount of time, before resolving to one of the stable
saturation states. Due to their function $V_{meta}$ must be in between their
regular HI and LO states, and, due to symmetry in the design, it is typically an
intermediate voltage level in the undefined range $V_{xx}$. With an
appropriately designed threshold of the subsequent stage this creeping behavior
can be transformed into a so-called \textit{late transition} where the output of
that stage shows a clean transition (i.e.\ fast crossing of the intermediate
levels) but only after metastability has resolved. However, with a single
threshold (i.e.\ without hysteresis) one also introduces the risk of glitches
\cite{PS13}.

Metastability is a very undesired phenomenon, as $V_{meta}$ may, beyond the
above-mentioned glitches, lead to different (``Byzantine'') interpretations by
input stages it supplies (as these will most likely have slightly different
thresholds), while a late transition can cause timing violations
downstream. Unfortunately, in general it can not be avoided completely.

Note that the above applies to bistable storage elements, whose metastable
behavior is already well researched -- we will have to revisit this for the
\gls{st}.

\subsection{Feedback Circuits}

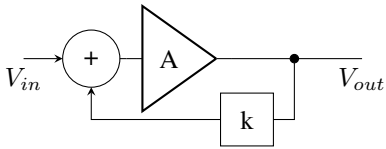
\begin{figure} [h]
  \centering
  \begin{tikzpicture}[>=stealth, scale=0.9, yscale=0.9]

  \node[circle, draw, outer sep=0, inner sep=5pt] (P) at (1,0) {+};
  \node[circ] (CONN) at (4,0) {};
  \node[draw, rectangle, outer sep=0, minimum width=20pt, minimum height=20pt] (k) at (3.3,-1) {k};
  \node[buffer] (B) at (2.3,0) {};
  \node at (2.15,-0.01) {A};
  
  \draw[->] (0,0) node [anchor=north] {$V_{in}$} -- (P);
  \draw[->] (k.west) -| (P);
  \draw (P.east) -- (B.in);
  \draw (B.out) -- (CONN);
  \draw (CONN) |- (k.east);
  \draw (CONN) -- ++(1,0) node [anchor=north] {$V_{out}$};

\end{tikzpicture}
  \caption{Basic structure of a feedback circuit }
  \label{fig:feedback}
\end{figure}

The arrangement shown in Fig.~\ref{fig:feedback} represents the fundamental
layout of a feedback circuit. A linear voltage amplifier with gain $A$ receives
as its input the sum of an external input voltage and its own output voltage
multiplied with a factor of $k$. Its (static) transfer characteristic can be
described by
\begin{equation}
G = \frac{V_{out}}{V_{in}} = \frac{-1} {k - \frac{1}{A}}
\label{equ:gain}
\end{equation}

In the case of $k<0$ we have \textit{negative feedback}. For the moment, let us
assume $A=\infty$. Then the arrangement operates as an amplifier with (positive)
gain $G$ of $-1/k$. For $k=-1$ we feed back the full output voltage and obtain
$G=1$, i.e.\ a voltage follower.  For $k=0$ we have no feedback, hence an
\textit{open loop}. This arrangement resembles the function of an ideal
discriminator whose output assumes the positive saturation voltage $M$ in case
$V_{in}>0$ and changes to the negative saturation\footnote{For simplicity of
  explanation we assume symmetric saturation voltages, i.e.\ +M and -M
  here. Although the quantitative results will differ in the asymmetric case,
  our reasoning and our basic conclusions will still hold.} $-M$ as soon as
$V_{in}<0$.

With $k>0$ we realize \textit{positive feedback}. Now every little change
$\varepsilon$ on the input produces a change on the output in the same direction
that gets fed back and thus further supports the original input change by being
added to $\varepsilon$. This self-supporting chain ultimately causes $V_{out}$
to run into positive or negative saturation.  In this situation the loop feeds
back a voltage of $V_{FB} = Mk$ (or $-Mk$, respectively) that must be
compensated in the summation by the input voltage, i.e.\ $V_{in}<-Mk$ (or
$V_{in}>Mk$, respectively) to move the output to the other direction, where it
again saturates. This resembles the function of a \emph{Schmitt-Trigger} with
hysteresis $V_{hyst} = V_H-V_L=2Mk$.

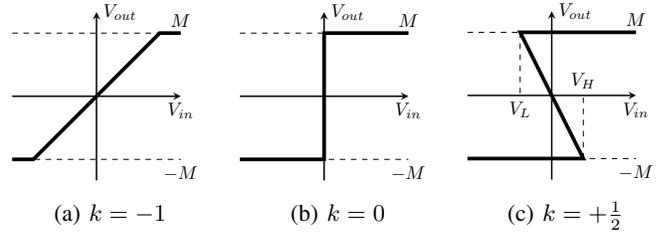
\begin{figure} [h]
  \centering
  \begin{subfigure}{0.3\linewidth}
    \scalebox{0.7}{\begin{tikzpicture}[>=stealth, scale=0.4]

  \def\lim{4}
  \def\valM{3}
  \def\kOne{3}
  \def\kHalf{1.5}
  
  \draw[line width=0.8pt,->] (-\lim,0) -- (\lim,0) node[anchor=north] {$V_{in}$};
  \draw[line width=0.8pt,->] (0,-\lim) -- (0,\lim) node[anchor=west] {$V_{out}$};  

  \draw[dashed] (-\lim,\valM) -- (\lim,\valM) node [anchor=south] {$M$};
  \draw[dashed] (-\lim,-\valM) -- (\lim,-\valM) node [anchor=north] {$-M$};

  \draw[black!100!white, line width=2pt] (-\lim,-\valM) -- (-\kOne,-\valM) --
  (\kOne,\valM) -- (\lim,\valM);



  
\end{tikzpicture}}
    \caption{$k=-1$}
  \end{subfigure}
  ~
  \begin{subfigure}{0.3\linewidth}
    \scalebox{0.7}{\begin{tikzpicture}[>=stealth, scale=0.4]

  \def\lim{4}
  \def\valM{3}
  \def\kOne{3}
  \def\kHalf{1.5}
  
  \draw[line width=0.8pt,->] (-\lim,0) -- (\lim,0) node[anchor=north] {$V_{in}$};
  \draw[line width=0.8pt,->] (0,-\lim) -- (0,\lim) node[anchor=west] {$V_{out}$};  

  \draw[dashed] (-\lim,\valM) -- (\lim,\valM) node [anchor=south] {$M$};
  \draw[dashed] (-\lim,-\valM) -- (\lim,-\valM) node [anchor=north] {$-M$};


  \draw[black!100!white, line width=2pt] (-\lim,-\valM) -- (0,-\valM) -- 
  (0,\valM) -- (\lim,\valM);


  
\end{tikzpicture}}
    \caption{$k=0$}
  \end{subfigure}
  ~
  \begin{subfigure}{0.3\linewidth}
    \scalebox{0.7}{\begin{tikzpicture}[>=stealth, scale=0.4]

  \def\lim{4}
  \def\valM{3}
  \def\kOne{3}
  \def\kHalf{1.5}
  
  \draw[line width=0.8pt,->] (-\lim,0) -- (\lim,0) node[anchor=north] {$V_{in}$};
  \draw[line width=0.8pt,->] (0,-\lim) -- (0,\lim) node[anchor=west] {$V_{out}$};  

  \draw[dashed] (-\lim,\valM) -- (\lim,\valM) node [anchor=south] {$M$};
  \draw[dashed] (-\lim,-\valM) -- (\lim,-\valM) node [anchor=north] {$-M$};



  \draw[black!100!white, line width=2pt] (-\lim,-\valM) -- (\kHalf,-\valM)
  -- (-\kHalf,\valM)  -- (\lim,\valM);

  \draw[line width=0.5pt, dashed] (\kHalf,-\valM) -- (\kHalf,0) node [anchor=south] {$V_H$} ;
  \draw[line width=0.5pt, dashed] (-\kHalf,\valM) -- (-\kHalf,0) node [anchor=north] {$V_L$};
  
\end{tikzpicture}}
    \caption{$k=+\frac{1}{2}$}
  \end{subfigure}
  \caption{Transfer characteristic of a feedback circuit }
  \label{fig:characteristics}
\end{figure}

Fig.~\ref{fig:characteristics} shows the characteristics $V_{out}$ over $V_{in}$
for different selections of $k$. We observe that for negative feedback we have a
unique mapping from $V_{in}$ to $V_{out}$, while for positive feedback $V_{out}$
depends on the current state for $V_L \leq V_{in} \leq V_H$, i.e.\ we have a
hysteresis behavior.  Note carefully that the saturation states are the only
``truly'' stable states of the \gls{st}.  The line described by
Eq.~\ref{equ:gain} describes the metastable states only. A very intuitive
explanation of this fact is that for a given input voltage
$V_L \leq V_{in} \leq V_H$ we can draw a vertical line to find the corresponding
steady-state values of $V_{out}$. This line has three intersections with the
characteristic, namely at the positive and negative saturation, as well as one
in between. Since we know that the saturation states represent truly stable
states, there must be a metastable state in between -- irrespective of the
implementation.  The transient behavior, i.e.\ the transition from one
saturation voltage to the opposite one, depends on the dynamic characteristics
of the circuit which are not considered in the basic model in
Fig.~\ref{fig:feedback}.

For a non-ideal amplifier with $A<\infty$ we obtain a reduction of the effective
$k$ by $\frac{1}{A}$ (see Eq~\ref{equ:gain}). In case of negative feedback this
reduces the overall gain accordingly, and in case of the \gls{st} it moves the
thresholds towards a reduced hysteresis. The borderline case of discriminator
operation now occurs for $k=\frac{1}{A}$. Apart from that shift in the value of
$G$ (that can be compensated by appropriate dimensioning), all qualitative
findings from above, however, remain the same. In Fig.~\ref{fig:characteristics}
we would, e.g., simply have to replace all instances of $k$ by
$k-\frac{1}{A}$. Furthermore, a reference voltage can be added to the feedback
path to create a hysteresis that is no more centered around $V_{in}=0$. Again,
while this shift obviously changes the quantitative results, the qualitative
findings still hold.

\subsection{Schmitt-Trigger Implementation}

A straightforward implementation of the principle from Fig.~\ref{fig:feedback}
is by means of an operational amplifier (OpAmp).  Since OpAmps usually have a
high gain, this implementation is close to the ideal case of
$A \rightarrow \infty$.

For negative feedback the feedback path is simply connected to the inverting
input, thus effectively realizing the negative sign.  For positive feedback the
non-inverting input of the OpAmp must be used, which leaves only the inverting
input for connecting $V_{in}$. This means that from the view of the input
voltage the function of the \gls{st} circuit is inverting.

In both cases a resistive voltage divider can establish
$|k|=\frac{R_B}{R_A+R_B}<1$, and the feedback path can be augmented with a
reference voltage source $V_R$ that will create a horizontal shift of the
characteristic. In case of positive feedback this results in a shift of the
threshold voltages by an amount of $(1-k)V_R$.

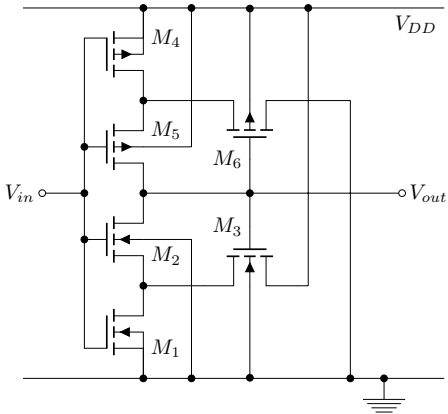
\begin{figure} [h]
  \centering
  \scalebox{0.8}{\begin{tikzpicture}[yscale=1]

  \draw
  (0,0) node[pigfete,yscale=1] (M4) {}
  (M4.D) node[pfet,yscale=1,anchor=S] (M5) {}
  (M5.D) node[nfet,anchor=D] (M2) {}
  (M2.S) node[nigfete,anchor=D] (M1) {}
  (M1.G) to [short,-*] (M2.G) to [short,-*] node[pos=0] (IN) {} (M5.G)  to [short] (M4.G)
  (IN) to [short,*-o] ++ (-0.7,0) node[anchor=east] (VIN) {$V_{in}$}
  (M4.D) to [short,*-] ++(1,0) node[pfet, xscale=1,rotate=90, anchor=S] (M6) {}
  (M1.D) to [short,*-] ++(1,0) node[nfet, rotate=270, anchor=S] (M3) {}
  (M2.D) to [short,*-o] ++(4.3,0) node[anchor=west] (VOUT) {$V_{out}$}
  (M1.S) to [short,*-] ++(-2,0)
  (M1.S) to [short,-*] ++(4,0) node [ground] (GND) {}
  (GND) to [short] ++ (1,0)
  (M4.S) to [short,*-] ++(-2,0)
  (M4.S) to [short] ++(5,0) node [anchor=north east] (VDD) {$V_{DD}$}
  (M3.G) to [short] node[pos=0,circ] {} (M6.G)
  (M5.B) to ++(0.8,0) to [short,-*] ([xshift=0.8cm]M4.S)
  (M6.B) to [short,-*] ++(0,1.53)
  (M2.B) to ++(0.8,0) to [short,-*] ([xshift=0.8cm]M1.S)
  (M3.B) to [short,-*] ++(0,-1.53)
  (M3.D) to ++(0.2,0) to [short,-*] ++(0,4.62)
  (M6.D) to ++(0.9,0) to [short,-*] ++(0,-4.62)
  (M2.B) node [anchor=north west] {$M_2$}
  (M1.B) node [anchor=north west] {$M_1$}
  (M4.B) node [anchor=south west] {$M_4$}
  (M5.B) node [anchor=south west] {$M_5$}
  (M3.G) node [anchor=east] {$M_3$}
  (M6.G) node [anchor=east] {$M_6$}
  ;
\end{tikzpicture}}
  \caption{Conventional CMOS \gls{st} implementation (from \cite{FB94})}
  \label{fig:CMOS_implementation}
\end{figure}

In digital CMOS logic circuits OpAmps are expensive to realize. Therefore a
different circuit structure has become common, namely a kind of extended
inverter stack with feedback from the output, as shown in
Fig.~\ref{fig:CMOS_implementation}.

There are different variations to this basic scheme, targeting at low supply
voltage \cite{SSK11} or adjustable thresholds \cite{Wang91,KGC08}.  In
\cite{FB94} a detailed mathematical analysis of this circuit is given.

\subsection{Metastability Model for the Schmitt-Trigger}
There has been (and sometimes still is) uncertainty whether or under which
circumstances a \gls{st} is prone to metastability, as its function of
classifying an input voltage (rather than storing data) appears combinational.
However, as Fig.~\ref{fig:characteristics} and the associated explanation in
Sec.~\ref{sec:metastability} show, \textit{every} positive feedback circuit
\textit{must} have a metastable state. Note, that this curve resembles a general
model of a positive feedback circuit without being limited to a specific
implementation.

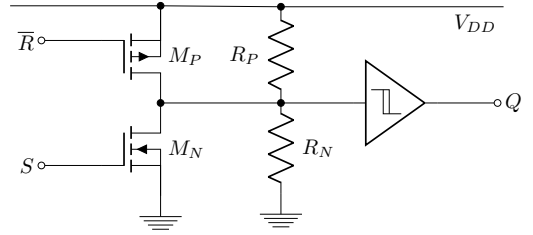
\begin{figure} [h]
  \centering
  \scalebox{0.8}{\begin{tikzpicture}

  \draw
  (0,0) node[pigfete,yscale=1] (MP) {}
  (MP.B) node[anchor=west] {$M_P$}
  (MP.D) node[nigfete,anchor=D] (MN) {}
  (MN.B) node[anchor=west] {$M_N$}
  (MP.S) to[short] ++(0,0.076) node[circ] (VDD1) {}
  (MN.S) node[ground] {}
  (MP.D) to[short,*-*] ++(2,0) node (CONN) {}
  (CONN) to[R=$R_P$] ++(0,1.6) node[circ] (VDD2) {}
  (CONN) to[R=$R_N$] ++(0,-1.5) node [ground] {}
  (MP.G) to[short,-o] ++(-1,0) node [anchor=east] {$\overline{R}$}
  (MN.G) to[short,-o] ++(-1,0) node [anchor=east] {$S$}
  (CONN) to[short] ++(1.2,0) node[buffer,anchor=in] (ST) {}
  (ST.out) to [short,-o] ++(1,0) node [anchor=west] {$Q$}
  (VDD1) to ++(-2.5,0)
  (VDD1) to (VDD2)
  (VDD2) to[short] ++(3.7,0) node [anchor=north east] {$V_{DD}$}
  ;

  \draw (3.5,-0.55) -| (3.8,-0.95);
  \draw (3.7,-0.55) |- (4.0,-0.95);

\end{tikzpicture}}
  \caption{RS-latch implementation based on a \gls{st}}
  \label{fig:ST_latch}
  \vspace{-0.27cm}
\end{figure}

Fig.~\ref{fig:ST_latch} shows how an RS-latch can be constructed from a
\gls{st}. This gives an intuition that the latter must have storage
capabilities. It also indicates that building a perfect \gls{st} is equivalent
to building a perfect RS-latch, which we know is impossible \cite{Mar81}.

\begin{figure} [h]
  \centering
  \scalebox{0.65}{\begin{tikzpicture}[scale=0.85, yscale=0.9]

  \draw
  (0,0) node[ocirc] (IN) {} to[short] ++(1,0) node[op amp, anchor=-] (opamp) {}
  (IN) node [anchor=east] {$V_{in}$}
  (opamp.out) to[R=$R_0$] ++(2.7,0) node[circ] (CONN2) {}
  (CONN2) to[C,l_=$C_0$] ++(0,2.2) node[ground, yscale=-1] {}
  (CONN2) to[short] ++(1.4,0) node[ocirc] {} node[right] {$V_{out}$}
  (CONN2) to[R=$R_A$] ++(0,-1.9) node[circ] (CONN3) {}
  (CONN3) to[R=$R_B$] ++(0,-1.9) node[ocirc] {} node[below] {$V_R$}
  (opamp.+) to ++(-1,0) |- (CONN3)
  ;
  
\end{tikzpicture}}
  \caption{Dynamic model of the \gls{st} inspired by Marino \cite{Marino77}}
  \label{fig:Marino_Circuit}
  \vspace{-0.05cm}
\end{figure}
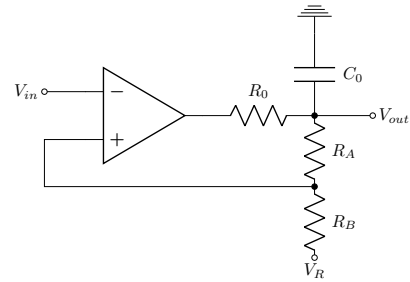

There has also been quite some debate on whether a \gls{st} can be applied to
construct a synchronizer flip flop that is immune against metastability (e.g.\
\cite{Worm77,Chaney79}). This discussion has been resolved in a paper by Marino
\cite{Marino77} in which he proposes a dynamic model for the \gls{st} circuit as
follows: He augments the OpAmp realization with a low-pass ($R_0 C_0$) at the
output to account for its dominant time constant, and thus approximates the
dynamic behavior in the model.  He assumes another RC element at the inverting
input and a reference voltage $V_R$ to obtain thresholds that are not bound to
be symmetric around 0.  A simplification of his model circuit used in this paper
is shown in Fig.~\ref{fig:Marino_Circuit}.  To be more general, we will,
compared to Marino's circuit, drop the input RC element, and we use more
intuitive names ($V_{in}$ for $V_2$ and $V_{out}$ for $V_q$).

\begin{figure} [h]
  \centering
  \scalebox{0.8}{\begin{tikzpicture}[>=stealth, font=\small, yscale=0.9]

  \def\lim{4}
  \def\limNeg{-1}
  \def\valM{3}
  \def\VP{4}
  \def\VN{2}
  
  \draw[->] (\limNeg,0) -- (\lim+2,0) node[anchor=north] {$V_{in}$};
  \draw[->] (0,-\lim) -- (0,\lim) node[anchor=north east] {$V_{out}$};  

  \draw
  (0.5,-\lim)[dashed] -- node[pos=0.82, anchor=east] (L1)
  {$V_{in}=\frac{R_AV_R+R_BV_{out}}{R_A+R_B}-\frac{M}{A}$} (4.5,\lim) 
  (1.5,-\lim)[dashed] -- node[pos=0.3, anchor=west] (L3)
  {$V_{in}=\frac{R_AV_R+R_BV_{out}}{R_A+R_B}+\frac{M}{A}$}(5.5,\lim) 
  ;
  
  \draw[line width=1.2pt]
  (\limNeg,\valM) -- node[anchor=south] {$V_{out} = \gamma_1$} (\VP,\valM)
  (\VN,-\valM) -- node[anchor=south] {$V_{out} = \gamma_3$}(\lim+2,-\valM)
  (1.6666,-\lim) -- node[fill=white, anchor=west, pos=0.6, xshift=0.1cm] (L2) {$V_{out}=\gamma_2$}(4.3333,\lim)
  ;

  \draw[->] (L1.south) -- ++(1.3,-0.3);
  \draw[->] ([xshift=0.2cm]L2.south west) -- ++(-0.3,-0.2);
  \draw[->] (L3.north) -- ++(-1.3,0.4);
  
  \draw[<-] (\limNeg,-3.8) --
  node [yshift=0.1cm, pos=0.5,fill=white, anchor=south, font=\footnotesize] {REGION 1}(0.6,-3.8);
  \draw[<->] (0.6,-3.8) --
  node [yshift=-0.1cm, pos=0.5,fill=white, anchor=north, font=\footnotesize] {REGION 2} (1.6,-3.8);
  \draw[->] (1.55,-3.8) --
  node [yshift=0.1cm, pos=0.5,fill=white, anchor=south, font=\footnotesize] {REGION 3}(\lim+2,-3.8);
  
\end{tikzpicture}}
  \caption{Phase diagram for the \gls{st} inspired by Marino \cite{Marino77}}
  \label{fig:Phase_Diagram}
  \vspace{-0.05cm}
\end{figure}
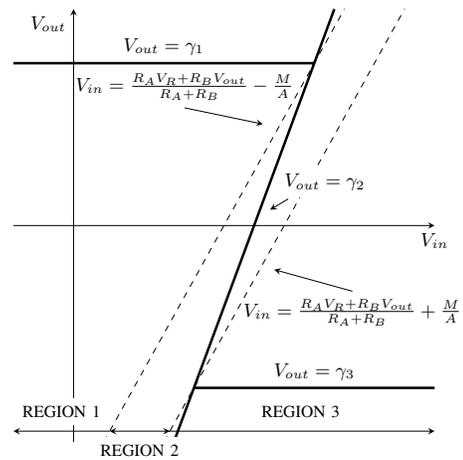

As the saturation requires separate treatment, his solution comprises three
regions, namely upper and lower saturation (Regions 1 and 3), as well as the
``linear region'' 2 between them.  Fig.~\ref{fig:Phase_Diagram} illustrates his
solution.  Note that, according to the implementation of the model circuit, this
diagram applies to an inverting \gls{st}.

The dashed lines in the figure represent the borders between the regions (the
corresponding equations are shown as well). They are derived by determining
those values for the OpAmp's differential input voltage for which it starts to
saturate. Ultimately, Marino obtains the following equations (for their detailed
derivation please refer to the original paper):

\begin{equation}
\label{eq:I_diff}
\text{Region~1:\hspace{1cm}}
\frac{dV_{out}}{dt} = V_{out}' = -\frac{1}{\tau_1} (V_{out} - \gamma_1)
\end{equation}
This yields a decaying exponential function with time constant
$\tau_1 \approx R_0 C_0$ that asymptotically approaches the truly stable rest
point $\gamma_1 \approx M$.

\begin{equation}
\label{eq:II_diff}
\text{Region~2:\hspace{1.4cm}}
\frac{dV_{out}}{dt} = V_{out}' = \frac{1}{\tau_2} (V_{out} - \gamma_2)
\end{equation}
This time we have a growing exponential function with time constant
$\tau_2 \approx \frac{R_0 C_0}{kA-1}$ that moves away from the metastable rest
point $\gamma_2 \approx \frac{V_{in}-(1-k)V_R}{k-\frac{1}{A}}$.  Note that this
rest point now depends on the input voltage.

\begin{equation}
\label{eq:III_diff}
\text{Region~3:\hspace{1cm}}
\frac{dV_{out}}{dt} = V_{out}' = -\frac{1}{\tau_3} (V_{out} - \gamma_3)
\end{equation}
Similar to Region~1 this yields a decaying exponential function with time constant 
that asymptotically approaches the truly stable rest point
$\gamma_3 = -\gamma_1 \approx -M$.

Marino uses this model to show that a \gls{st} can neither be used to build a
perfect inertial delay element nor a perfect synchronizer. Although his model
clearly shows that a \gls{st} may indeed become metastable, in his argumentation
he is mainly concerned with the question of whether it can be driven to produce
runt pulses or glitches at its output. He shows that this is indeed possible,
even if restrictions for the input are applied.  He does, however, not consider
other metastability effects like transition delay or constant output voltage and
under which circumstances these occur. Similarly, Nystr\"om and Martin
\cite{Nystroem} as well as Greenstreet \cite{Greenstreet} limit their
discussions to the special case of monotonic input voltage only.

In this paper we build on Marino's model, but extend the scope towards typical
use cases, validate the model for a modern CMOS implementation, provide a more
general treatment of the metastable behavior of a \gls{st}, and give a deeper
analysis of the case of monotonic input voltage than in \cite{Nystroem,
  Greenstreet}.

\section{Analysis of Metastable Behavior}
\label{sec:analysis}

\subsection{Peculiarities in the Schmitt-Trigger's metastable behavior}
\label{sec:peculiar}
Fig.~\ref{fig:Phase_Diagram} suggests that a \gls{st} can assume different
metastable points $V_{meta}$; in fact along the whole line $\gamma_2$. This is
substantially different from what is known for typical bistable storage
elements, whose (internal storage cell's) metastable output voltage is confined
to a single value in the $V_{xx}$ range.  The metastable behavior of a latch
cell has been first modeled by Veendrick \cite{Vee80}, and his analysis forms
the theoretical foundation of what we know about metastable behavior of bistable
storage elements today. So let us compare Marino's model with that used by
Veendrick to spot the differences. In both models a linear amplifier is
employed, and its dynamic behavior is approximated by a first order low pass. So
not surprisingly the solutions are exponential functions in both cases and hence
similar.  However there are two important differences:
\begin{enumerate}
\item For his latch circuit Veendrick assumes the input to be decoupled (opaque state) and just studies the homogeneous behavior, while Marino, for his \gls{st} model needs to leave the input voltage connected all the time. As a result, Marino's solution shows a dependence of the metastable rest point on the input voltage in Region~2, rather than just a single metastable point.
\item As a consequence of (1), Veendrick could concentrate his analysis on the proximity of the metastable point, while Marino had to consider the whole operating range and therefore needed the case separation.
\end{enumerate}
So (1) gives us an intuitive confirmation why the \gls{st} has a whole range of metastable points. Conceptually, this appears to be due to the fact that, having only one input, the \gls{st} derives its trigger for the state change from the amplitude of this single signal, making a constant observation necessary, while all bistable storage elements have two inputs and can hence decouple either of them temporarily.

\subsection{Regular operation}

We will base most of our reasoning on the phase diagram
(Fig.~\ref{fig:Phase_Diagram}). So let us first observe the normal operation of
the \gls{st} there: We start in the positive saturation in Region~1 (for the
rest of the paper we will always consider the positive saturation as a starting
point, while due to symmetry, equivalent arguments can be given for starting in
the negative saturation) As we increase $V_{in}$ we move along $\gamma_1$ until
we reach $V_H$. Up to this point we have no freedom in choosing the trajectory,
and the shape of $V_{in}$ is irrelevant.  Only after crossing $V_H$ the \gls{st}
will leave the stable state and start moving towards the negative
saturation. During this phase -- and only then -- we have the opportunity to
manipulate the trajectory and force the \gls{st} back to the initial state, or
maneuver it into a metastable state. Here the shape of $V_{in}$ matters a
lot. We will investigate more details on that later. Once in the negative
saturation, the same procedure starts over in the other direction.

\subsection{Monotonic input}
\label{sec:monotonic}

Let us again start on some point along $\gamma_1$. Exceeding $V_H$ then implies
a positive slope of $V_{in}$, and all trajectories reachable with a monotonic
$V_{in}$ are hence within the half plane $V_{in} > V_H$ where there is no
metastable point (recall that the latter are all located on $\gamma_2$). In fact
$V_{in}$ need not even be monotonic, as long as it does not fall back to below
$V_H$.

\begin{figure} [t]
  \centering
  \includegraphics{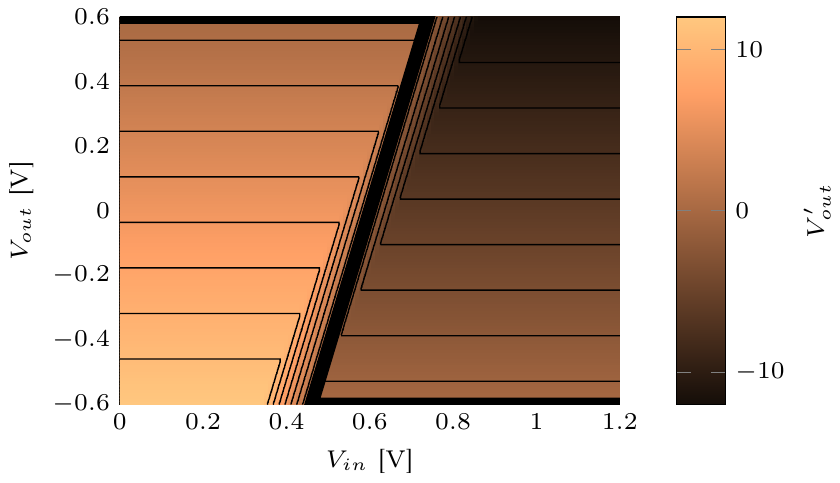}
  \caption{Derivative of the output voltage over the phase diagram}
  \label{fig:3Dgradients}
  \vspace{-0.3cm}
\end{figure}

Fig.~\ref{fig:3Dgradients} shows the first derivative $dV_{out}/dt$ over the
phase diagram according to Eq.~\ref{eq:I_diff} to \ref{eq:III_diff}. We have
chosen $A=10$ for this plot, which is way too low for a typical OpAmp, but for
higher values Region~2 would be hard to recognize (its width is just
$\frac{2M}{A}$). $V_{out}'$ represents the speed at which the trajectory is
pulled upward ($V_{out}'>0$) or downward ($V_{out}'<0$) by the internal dynamics
of the circuit.  We observe that if, starting from positive saturation, we apply
a step function to move to an operating point very close to but above the
threshold, say $(V_{in},V_{out})=(V_H+\varepsilon , \gamma_1)$ the downward
speed $V_{out}'$ is close to zero, so $V_{out}$ will initially change very
slowly. This suggests we obtain a slow output transition.
To determine the duration of this transition, let us first assume our step input
takes us right into Region~3, i.e.\ $\varepsilon > \frac{2M}{A}$. Then for
constant $V_{in}=V_H+\varepsilon$, Eq.~\ref{eq:III_diff} predicts a decaying
exponential function from $\gamma_1$ towards $\gamma_3$ according to

\begin{equation}
V_{out}(t) = (\gamma_1-\gamma_3) \cdot e^\frac{-t}{\tau_3} + \gamma_3
\label{eq:III_solution}
\end{equation}

Now we assume a threshold $V_{th}$ for the subsequent stage to recognize $V_{out}$ as being LO with $V_{th}= \gamma_3+\sigma\cdot(\gamma_1-\gamma_3)$, with $0<\sigma<1$ giving the proportion of the swing that $V_{th}$ is apart from the final value (that is reached asymptotically). This value will be reached with a delay of

\begin{equation}
D_\text{\em III} = \tau_3 \cdot ln\left (\frac{1}{\sigma}\right )
\label{eq:D_III}
\end{equation}

\noindent after having applied the input step. Note that, as long as we remain
within Region~3, this value is independent of $V_{in}$ (and hence $\varepsilon$)
and therefore stays the same, even if we apply larger steps. It is the minimum
switching time of the \gls{st}.

For $\varepsilon\in[0,\frac{2M}{A}]$ we start the trajectory in Region~2. Again
with constant $V_{in}$ it will move downward and cross the boundary to Region~3
at some point. Up to that point $V_{out}$ will follow a growing exponential
function according to

\begin{equation}
V_{out}(t) = - \frac{\varepsilon+\frac{\gamma_1-M}{A}}{k-\frac{1}{A}} \cdot e^{\frac{t}{\tau_2}}
+\frac{\gamma_1 k-\frac{M}{A}+\varepsilon}{k-\frac{1}{A}}
\label{eq:II_solution}
\end{equation}
\noindent and the time needed for the trajectory to move through Region~2 becomes approximately

\begin{equation}
D_\text{\em II} = \tau_2 \cdot ln \left(\frac{2M}{A\varepsilon} \right ).
\label{eq:D_II}
\end{equation}

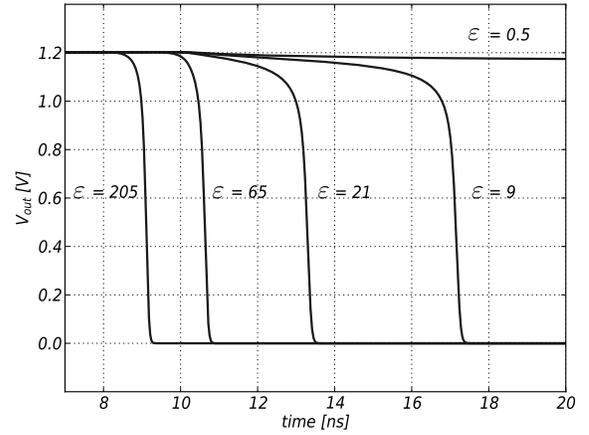
\begin{figure}[h]
  \centering
  \includegraphics[trim=0 0 0 12mm, clip, width=0.97\linewidth]{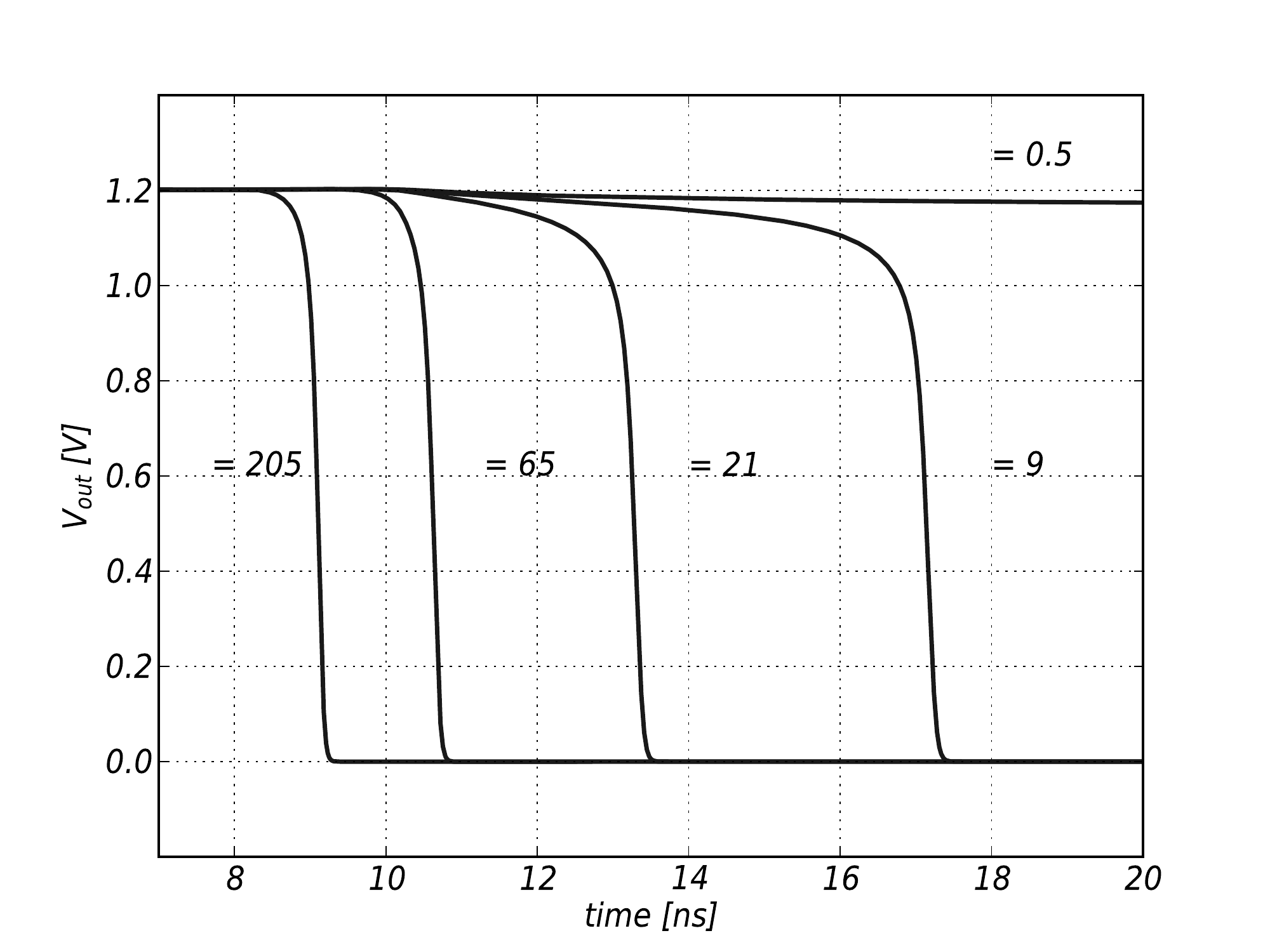}
  \begin{tikzpicture}[remember picture, overlay]
    \node at (-7.45,3.3) {$\varepsilon$};
    \node at (-5.6,3.3) {$\varepsilon$};
    \node at (-4.2,3.3) {$\varepsilon$};
    \node at (-2.15,3.3) {$\varepsilon$};
    \node at (-2.2,5.4) {$\varepsilon$};
  \end{tikzpicture}
  \caption{Falling output transitions for different $\varepsilon$ in mV}
  \label{fig:ST_switch_waveform}
  \vspace{-0.3cm}
\end{figure}

At the region boundary the decaying function from Eq.~\ref{eq:III_solution} will take over. Fig.~\ref{fig:ST_switch_waveform} shows a simulation result (for details on the setup see Section~\ref{sec:evaluation}) that illustrates the situation.

The transition time is the sum of $D=D_\text{\em II}+D_\text{\em III}$ (with a
small error due to $D_\text{\em III}$ actually being valid for the full swing).

The simulation results for the CMOS implementation shown in
Fig.~\ref{fig:ST_switch_waveform} confirm that the simplified OpAmp model does a
good job in predicting the behavior. In particular one can verify that
$D_\text{\em II}$ dominates, especially for small $\varepsilon$.  It is
interesting to note that using Veendrick's model \cite{Vee80} for calculating
the required resolution time $D_{meta}$ of a latch from a metastable state
yields, similar to our result, a $D_{meta}$ proportional to the metastability
time constant $\tau_C$ and to $ln(\frac{1}{V_\Delta})$ with $V_\Delta$ being the
initial voltage disparity.

The $\gamma_n$ lines are by definition the only places where $V_{out}'$ becomes
zero. So, due to the continuity of $V_{out}'$ proven in \cite{Marino77}, the
sign of $V_{out}'$ stays the same as long as we do not cross a $\gamma_n$
line. This can also be verified in Fig.~\ref{fig:3Dgradients}. As a consequence
we have a strictly decreasing $V_{out}$ in the above cases of $\varepsilon > 0$,
even if the start may be arbitrarily slow.  Although this can be regarded as
what is normally called a late transition in the context of bistable elements,
we have a fundamentally different metastable behavior in the \gls{st}: This late
transition is due to resolution of a metastable state that is associated with a
clean HI level. In contrast, for bistable storage elements the metastable state
is necessarily associated with resting at an intermediate voltage in the range
$V_{xx}$ between the element's clean HI and LO outputs, and the late transition
is a secondary effect caused by applying a high or low threshold \cite{PS13}.

\subsection{Producing a constant output voltage}
\label{sec:constant}

Let us now study the possibility of driving the \gls{st} into an arbitrary
metastable state: Assume we start again on $\gamma_1$.  Once $V_{in}$ exceeds
$V_H$ our operating point is right of $\gamma_2$, and the internal dynamics of
the \gls{st} is moving us downward ($V_{out}'<0$). In order to reach a
metastable operating point on $\gamma_2$ we need to reduce $V_{in}$ fast enough
to make the trajectory intersect with $\gamma_2$ before the negative saturation
is reached. Due to the inclination of $\gamma_2$ the amount by which we have to
reduce $V_{in}$ grows as $V_{out}$ moves downward.  In addition, $V_{out}'<0$
becomes larger with the operating point's distance from $\gamma_2$. So once that
distance is large, it takes a highly dynamic change in $V_{in}$ to reach a
metastable point.  Contrariwise, when staying close to $\gamma_2$ right from the
start, $V_{out}'$ can be kept as small as desired, leaving enough time for an
arbitrarily slow change in $V_{in}$ to reach a metastable point on $\gamma_2$ at
any desired intersection point.

As can be seen in Fig.~\ref{fig:Phase_Diagram}, $\gamma_2$ provides a one-to-one
mapping between $V_{in}$ and $V_{out}$. So with an appropriate choice of the
final value of $V_{in}$ (i.e.\ once having the threshold crossings
accomplished), any value of $V_{out}$ can be selected.  Notice that this
property allows us to freely select the metastable output voltage of the
\gls{st} based latch sketched in Fig.~\ref{fig:ST_latch} by proper adjustment of
the voltage divider.

\subsection{Creating an arbitrary output shape}
\label{sec:arbitrary}

In principle, by appropriately navigating in the phase diagram one can obtain
any desired shape of $V_{out}$: For every current value of $V_{out}$ an
appropriate $V_{in}$ can be applied to obtain the desired gradient $V_{out}'$
(by crossing $\gamma_2$ even the sign can be changed). However, with a limited
range of $V_{in}$ only a limited range of $V_{out}'$ can be covered (see
Fig.~\ref{fig:3Dgradients}); in other words, the dynamics of $V_{out}$ is
naturally limited by the system dynamics. The second limitation is the dynamics
of $V_{in}$.  Assume an operating point with a horizontal distance $X$ and
vertical distance of $Y=\alpha \cdot X$ from $\gamma_2$, with
$\alpha \approx \frac{1}{k-\frac{1}{A}}$ being the slope of the
latter. According to Eq.~\ref{eq:II_diff} $V_{out}'$ has a value of
$\frac{Y}{\tau_2}$ at this point. Moving the trajectory closer towards
$\gamma_2$ takes a $V_{in}'$ larger (in absolute value) than
$\frac{V_{out}'}{\alpha}$. So for a given maximum gradient $\hat{V}_{in}'$, we
obtain a maximum allowed horizontal distance from $\gamma_2$ of
$|X|<\tau_2 \cdot |\hat{V}_{in}'|$.  Once the operating point leaves this
corridor around $\gamma_2$, there is no way of preventing the trajectory from
approaching the saturation of $V_{out}$ in a monotonic trace (For a more
elaborate and formal treatment see \cite{Marino77}).

\section{Evaluation}
\label{sec:evaluation}

\subsection{Setup and characteristic}

\begin{figure} [h]
  \vspace{-0.3cm}
  \centering
  \includegraphics[trim=0 2mm 0 12mm,clip,width=0.97\linewidth]{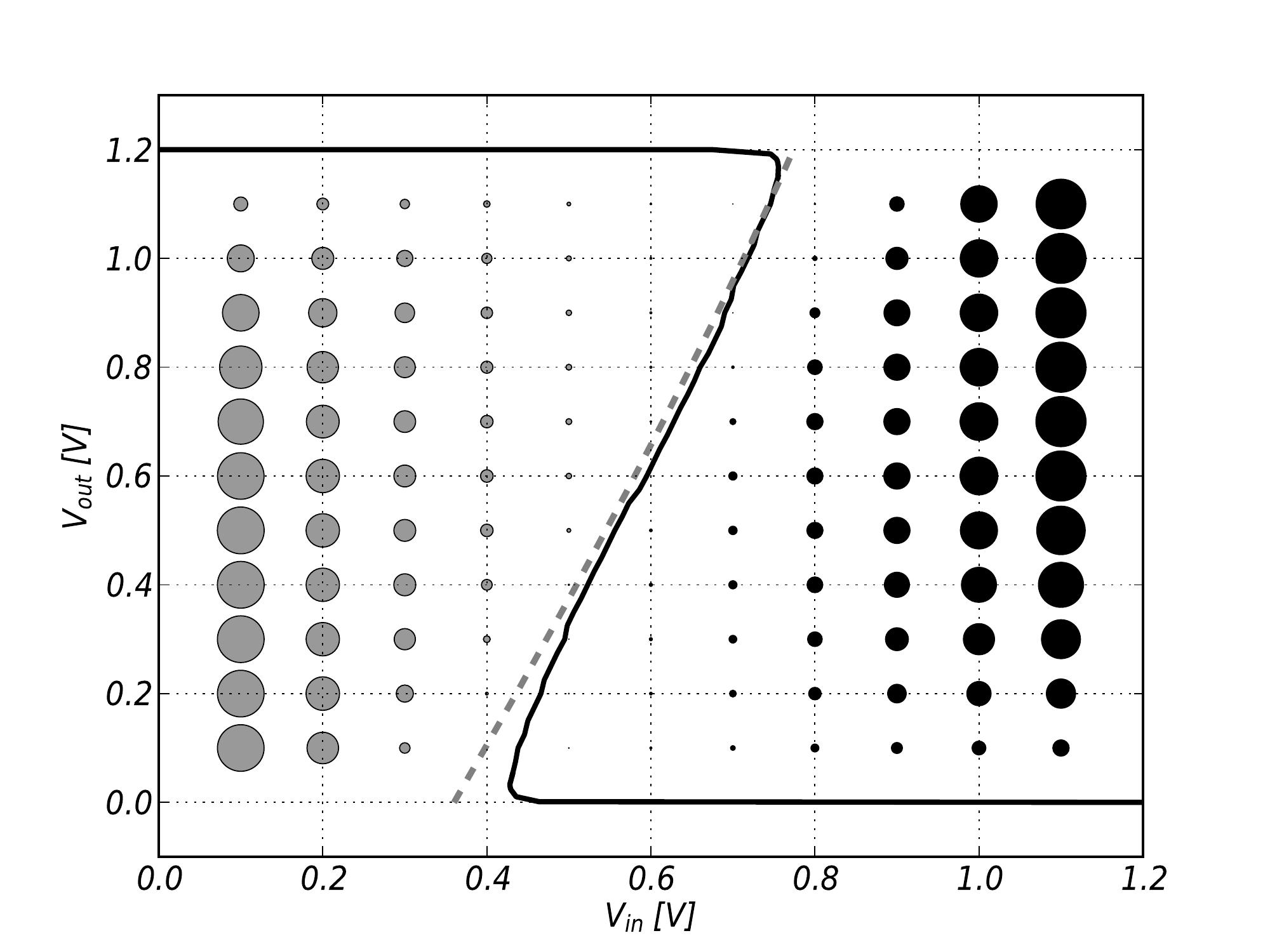}
  \caption{Derivative of the output voltage over the phase diagram}
  \label{fig:3Dgradients_measured}
  \vspace{-0.3cm}
\end{figure}

To validate our analyses we implemented \gls{st}s based on an ideal OpAmp, which
matched the theoretical model perfectly, a commercial OpAmp (Type EL5165), which
showed only minor deviations, and the CMOS circuit from
Fig.~\ref{fig:CMOS_implementation} in HSPICE.  As the latter is substantially
different from Fig.~\ref{fig:Marino_Circuit} we wanted to investigate whether
Marino's model sufficiently covers its behavior.  It was implemented using
transistor parameters of a standard inverter cell from an industrial 65\,nm
technology library, whereat despite a $2$\,fF output load no interconnect
parasitics were considered.  The resulting input-to-output characteristic is
shown in Fig.~\ref{fig:3Dgradients_measured}.  It matches the theoretical model
(Fig.~\ref{fig:Phase_Diagram}) well, however $\gamma_2$ turns out to be not
straight but shows an increased slope at the ends.  It was determined point by
point, in each case starting a transient analysis with a preset pair of
$\mathring{V}_{out}$ and $\tilde{V}_{in}$. By sweeping the value of
$\tilde{V}_{in}$ we determined the matching $\mathring{V}_{in}$ for which the
transient analysis showed stable behavior.  The dots in the figure represent
$V_{out}'$, with gray dots for positive values and black dots for negative ones,
and with large dot size indicating a large value. The large ``corridor'' around
$\gamma_2$ points to a wide Region~2 and hence a low gain $A$.  In addition, we
observe a dependence of $V_{out}'$ on $\gamma_2$ in the upper right and lower
left corner that is not present in the ideal model in
Fig.~\ref{fig:3Dgradients}.  The qualitative results from
Section~\ref{sec:analysis}, however, only require $V_{out}'$ to be consistently
positive (negative) on the left (right) side of $\gamma_2$ and continuous, so
they still hold.

We also analyzed the CMOS circuit using transistor equations to derive an
analytical expression for $\gamma_2$ (dashed line in
Fig.~\ref{fig:Phase_Diagram}).  By searching for equilibrium states, i.e.\ where
a constant input leads to a constant output voltage, an explicit formula
$V_{out}(V_{in})$ could be derived, assuming transistors $M_1$ and $M_4$ operate
in their linear region and all others in the saturation one. Unfortunately this
assumption is only valid in the middle of the metastable region; at the edges
the transistors $M_2$ and $M_3$ respectively $M_5$ and $M_6$ start to enter
their linear operation region. For that reason the analytic expression, while
matching with Marino's OpAmp model, does not fit well to the real curve near
$V_H$ and $V_L$.

\subsection{Evaluation of the scenarios from Section~\ref{sec:analysis}}

\begin{figure} [h]
  \centering
\includegraphics[trim=0 2 0 12mm,clip,width=0.97\linewidth]{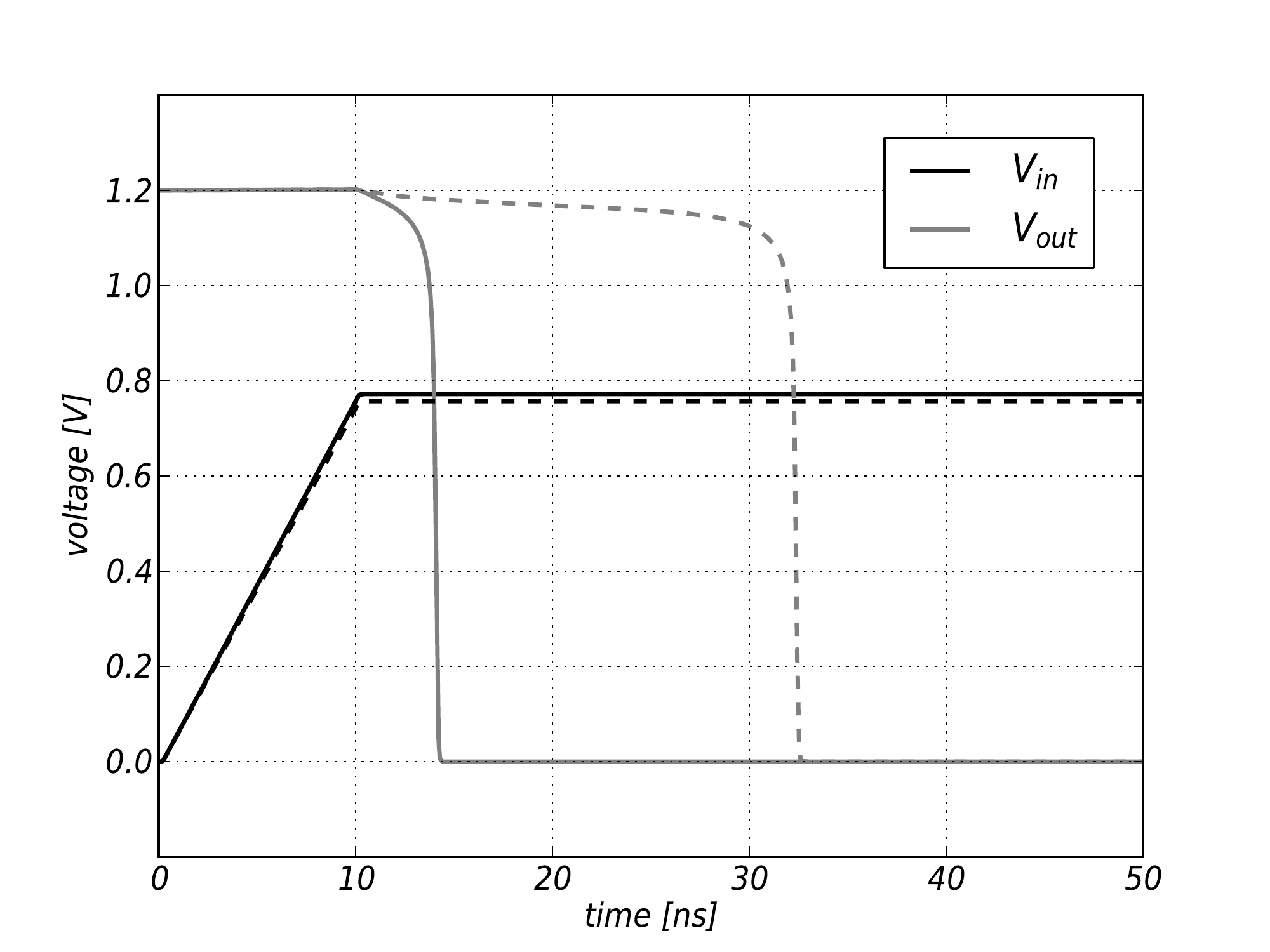}
\caption{Late transitions caused by ramp input going slightly above $V_H$}
\label{fig:late_transition}
\vspace{-0.5cm}
\end{figure}

Our claim was that monotonic input signals will always lead to strictly
monotonic outputs.  In the simulation shown in Fig.~\ref{fig:late_transition} we
verified the worst case by applying a ramp input stopping at a constant input
voltage close to $V_H$, i.e.\ $V_H+\varepsilon$ (dark lines).  One can clearly
see that in both cases the output transitions are very steep (all with about the
same transition time) but, as theory predicts, their delay varies significantly
even for small changes in $\varepsilon$.

Fig.~\ref{fig:delay_epsilon} illustrates the observed dependence of the output
delay on $\varepsilon$ in a more global scope. This nicely confirms
Eq.~\ref{eq:D_II}.

\begin{figure*}[ht!]
  \centering
  \begin{subfigure}{0.7\linewidth}
    \includegraphics[trim=0 0mm 0 2mm,clip, width=1\linewidth]{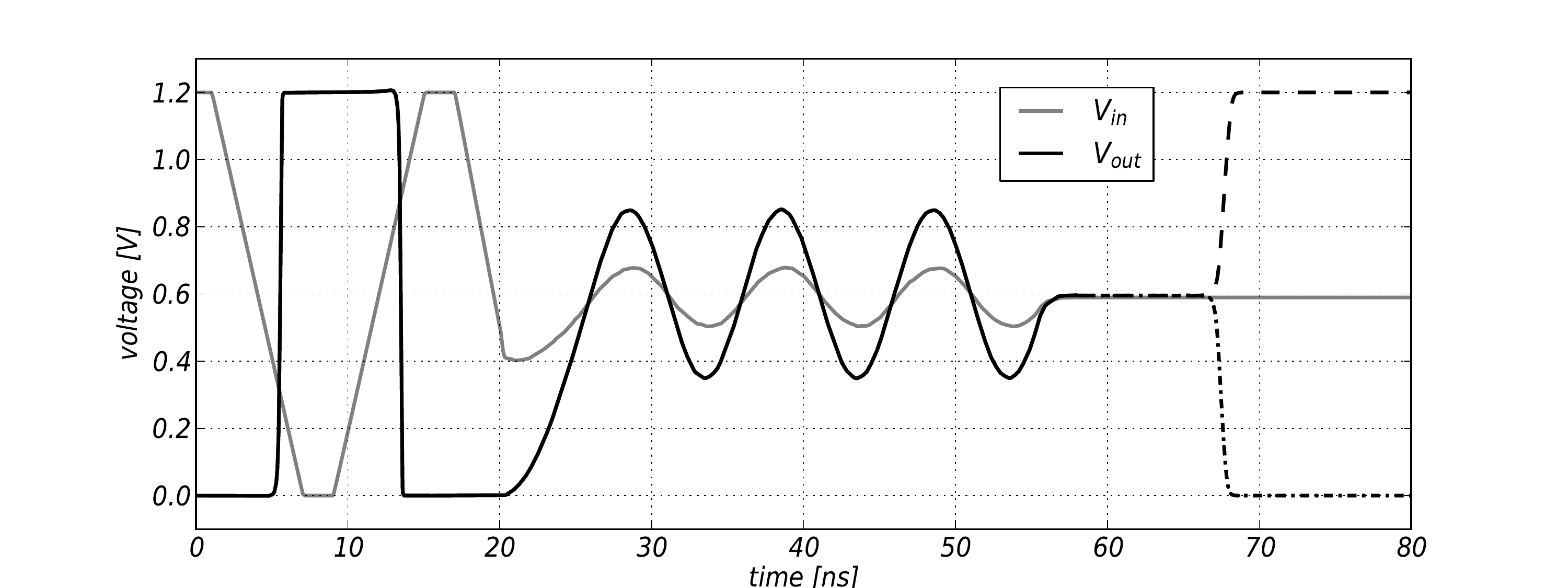}
  \end{subfigure}
  ~ \hspace{-0.5cm}
  \begin{subfigure}{0.29\linewidth}
    \includegraphics[trim=0 0mm 0 2mm,clip, width=1\linewidth]{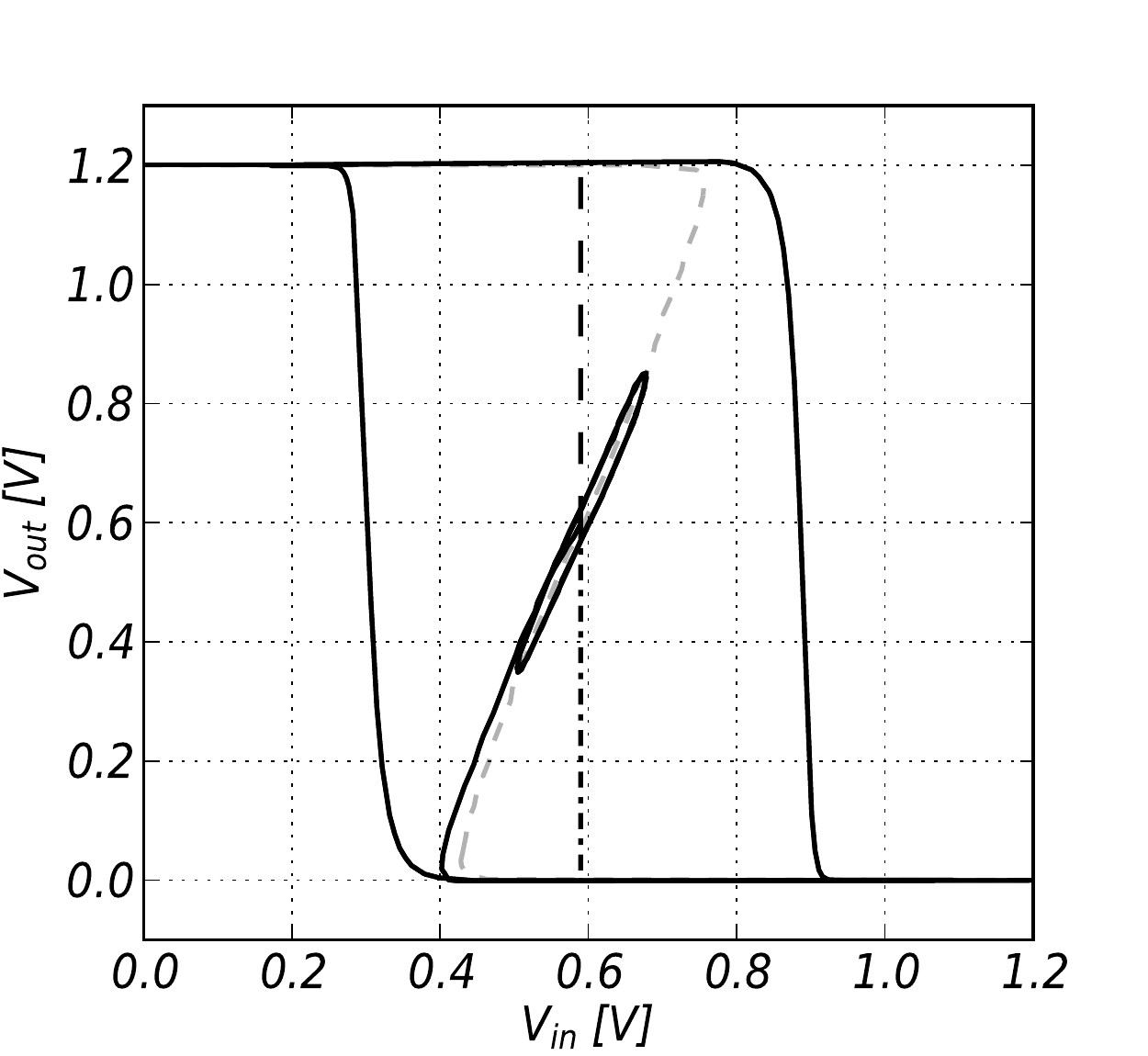}
  \end{subfigure}
  \caption{Simulation trace and phase diagram of an S/T driven to produce regular transitions, an arbitrary (here sine) waveform and to enter and resolve deep metastability}
  \label{fig:simTrace}
  \vspace{-0.5cm}
\end{figure*}

Fig.~\ref{fig:simTrace} shows that it is indeed possible to force the \gls{st}
to output arbitrary waveforms by means of non-monotonic inputs. In the first
part, the figure shows regular operation to demonstrate the dynamics of the
\gls{st} as well as its thresholds. Starting at 20\,ns, the \gls{st} is driven
to output a 100\,MHz sine with 0.5\,V swing. Note that this requires keeping the
\gls{st} metastable.  Finally, the simulated \gls{st} is driven into deep
metastablity with the input being constant from 58\,ns simulation time. Here,
the results of two simulations can be seen. In the first, metastability resolves
to $V_{DD}$, in the second, it resolves to \emph{GND}.

In the phase plot, it can be seen that the generation of the slow (w.r.t. its
regular switching speed) sine required to keep the output close to the
$\gamma_2$ line. The resolution of the metastable state can also be seen as
vertical line segments at $V_{in}\approx 0.6\,V$.

This verifies the predictions from Sections~\ref{sec:constant} and
\ref{sec:arbitrary}.
A constant output voltage can either be generated as an arbitrary waveform by
actively controlling the input, or by forcing the \gls{st} into perfect
metastability.  As before, small changes at $V_{in}$ lead to huge variations in
the time progress of $V_{out}$. The two output traces shown in the figure
correspond to input traces only deviating in their final stable voltage by less
than $0.1\, \mu \text{V}$ (not distinguishable in the figure). Clearly, if the
appropriate voltage is set with a sufficient precision, it can take an arbitrary
time for the metastability to resolve.

Nevertheless, in these simulations we experienced that it takes an
\textit{extremely} precise control of the voltage (nV) in order to get close
enough to $\gamma_2$ such that slow inputs still create visible metastability
effects, as theory would predict.

\section{Practical Use Cases}
\label{sec:usecases}

\subsection{Handling of intermediate voltages}
Often a \gls{st} is applied as a means for converting the intermediate voltage
$V_{meta}$ produced by a metastable binary storage element into a clean HI or
LO, like e.g.\ in \cite{PS13}. As we have seen in our analysis in
Section~\ref{sec:monotonic} this will actually work under two important
conditions:

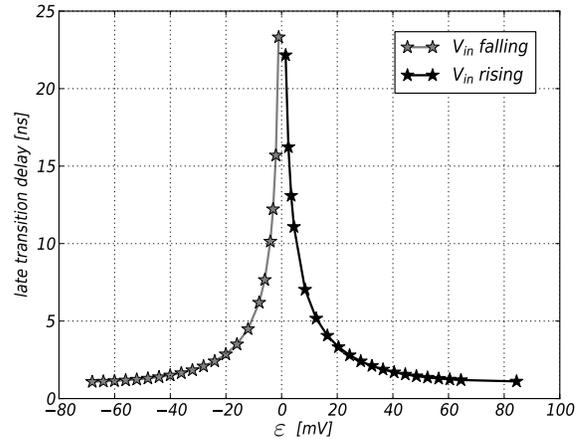
\begin{figure} [t]
\vspace{-0.1cm}
  \centering
  \includegraphics[trim=0 0 0 10mm,clip,width=0.97\linewidth]{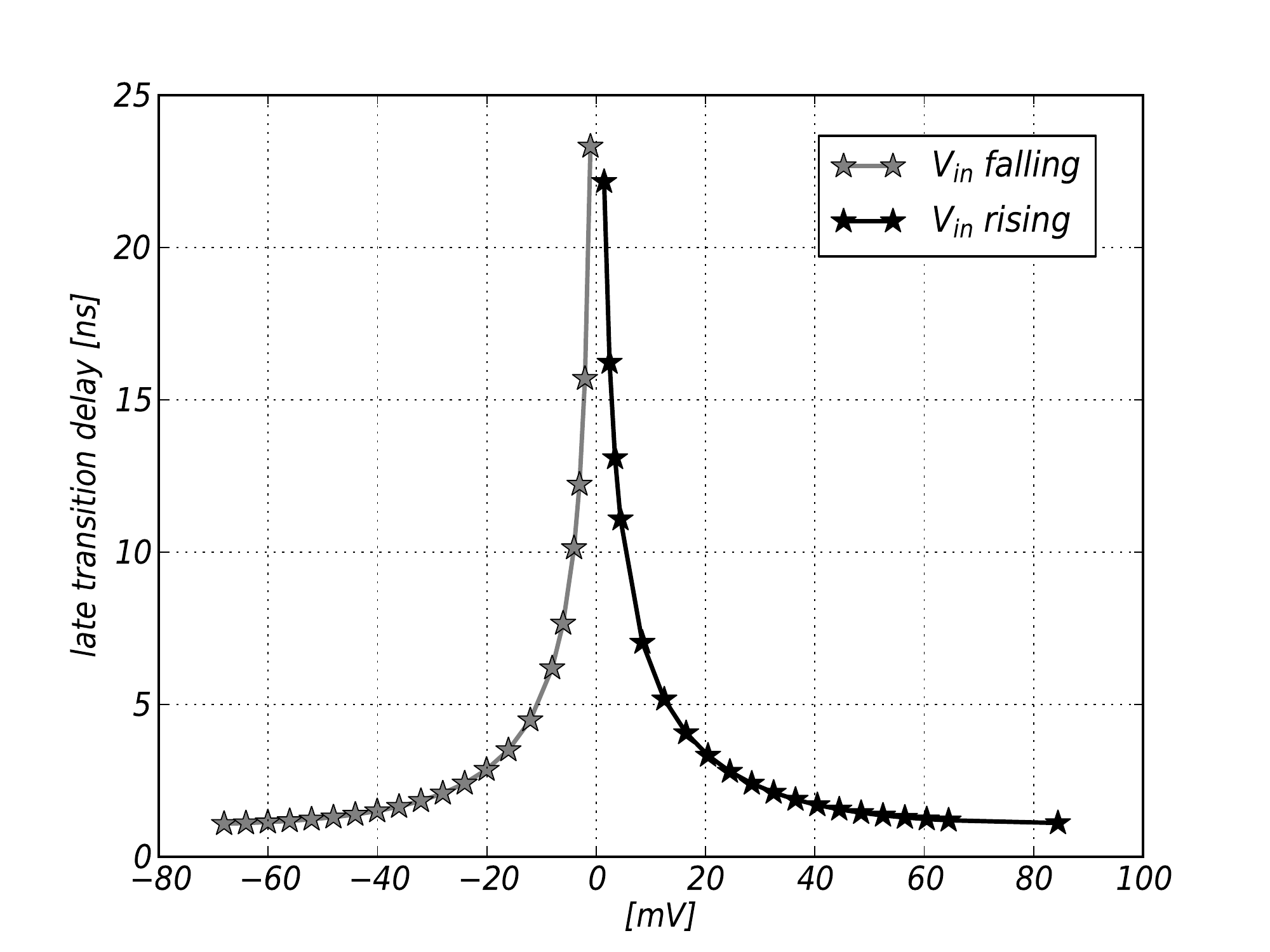}
  \begin{tikzpicture}[remember picture, overlay]
    \node at (-4.7,0.22) {$\varepsilon$};
\end{tikzpicture}
\caption{Observed dependence of output delay on $\varepsilon$}
\label{fig:delay_epsilon}
\vspace{-0.4cm}
\end{figure}

\textit{(1) The input of the \gls{st} must indeed be monotonic}, at least in the
proximity of the thresholds. This can be easily accomplished in a typical
setting, where the (single!) intermediate output voltage $V_{meta}$ is near the
middle of the supply range. With thresholds chosen in appropriate distance from
$V_{meta}$ one can ensure that these are crossed only when metastability is
already resolving, i.e.\ with an increasing exponential function that is
strictly monotonic (for details see \cite{Vee80,KC87}). However, care must be
taken that it is indeed the \gls{st} that decides upon the classification of
$V_{meta}$. As soon as any other stage (decoupling buffer, e.g.) is in between
the metastability-producing element and the \gls{st}, that element's (single!)
input threshold will typically classify $V_{meta}$ in an undesired way. More
specifically, glitches can be produced \cite{PS13}, with the \gls{st} having no
chance to mitigate these.

\textit{(2) A delay introduced by the \gls{st} must be accommodated in the
  timing of the subsequent logic.} With properly selected thresholds as outlined
above we can assume steep input transitions, so the \gls{st} will not by itself
introduce the arbitrary resolution delay discussed in
Section~\ref{sec:monotonic}. Still it may take an unbounded time until the
metastability of the bistable storage element resolves, during which the
\gls{st} observes a constant $V_{meta}$ at its input. As its threshold is
crossed only after that, the \gls{st} appears to produce a late transition. This
is actually an intended behavior, useful for handling metastability in a value
safe system, like a speed-independent design \cite{PS13}.

Essentially, (2) is the reason why Chaney \cite{Chaney79} and Kleeman et
al. \cite{KC87} correctly state that the use of \gls{st}s is not beneficial for
avoiding metastability in a synchronizer and even degrades the
performance. Marino \cite{Marino77}, on the other hand, was concerned with
inputs not limited to monotonic slope. Therefore his conclusion was, similarly,
that the \gls{st} is not useful in avoiding metastability. As we have laid out,
however, for the special application of filtering of intermediate voltages from
a metastable bistable storage element in value safe environments, the \gls{st}
can be safely applied without any residual risk of metastability.

\subsection{Slow inputs}
\label{sec:slow_inputs}
It is sometimes hoped \cite{web} that limiting the dynamics of the input signal
can prevent the \gls{st} from getting metastable. The intuition is that the
\gls{st} will have accomplished its state change before a (slow) change in the
input voltage has had a chance to move the trajectory towards a metastable
point.  Our analysis in Section~\ref{sec:arbitrary} has re-confirmed Marino's
result that one can always find a corridor around $\gamma_2$ small enough to
allow an appropriately controlled $V_{in}$ to still reach a metastable point, no
matter how restricted its dynamics ($\hat{V}_{in}'$) may be. However, as our
simulation experiments showed, it takes an extremely precise control of $V_{in}$
to remain in a sufficiently narrow corridor. So while limiting $V_{in}'$ cannot
safely rule out metastability of the \gls{st}, it \textit{does} aid in making
metastability less probable.

\subsection{Handling slow monotonic inputs}

We have given evidence in Section~\ref{sec:monotonic} that a \gls{st} can map
arbitrarily slow monotonic inputs to steep, practically full-swing transitions.
However, metastability can still occur and cause a seemingy sporadic transition
during a period with an unchanging input voltage.

One example of such an application is the \gls{st} D-latch implementation from
\cite{Greenstreet}, where the application requires handling glitches on the
enable input.  The input stack is a tri-state inverter that propagates the data
input when the latch enable is high, and has a floating output (assumed
constant) else.  The resulting monotonic signal is fed into a \gls{st}.  The
author correctly recognizes that even in presence of glitches on the enable
input, the \gls{st} would always correctly output steep transitions, albeit with
an arbitrary delay.

Another example is the integrator used in the synchronizer and clock to
handshake circuits in \cite{Nystroem}.  Here a precharged high signal is driven
low (or vice versa) depending on the state of an external, unstable input.  It
is also correctly argued that a \gls{st} converts these monotonic inputs to
steep transitions, however, the possibility that a signal driven slightly beyond
the \gls{st}'s threshold and left at that constant voltage may cause arbitrarily
delayed output transitions, is not further pursued.  The subsequent circuits,
being delay insensitive, can tolerate such delayed transitions, however one
should be aware of the possibility for such timing variations.

\section{Conclusion}
\label{sec:conclusion}

We have revisited existing results on \gls{st} metastability, most notably those
from Marino \cite{Marino77}, and extended them to elaborate a general
understanding of this effect and give well founded answers to a couple of
practical questions. In this sense our key contributions are to clearly pinpoint
the differences between \gls{st} metastability and that of bistable storage
elements, to provide simulation results from a realistic CMOS implementation
that back up theoretical results (shape of characteristic, $V_{out}'$ over the
phase diagram), to elaborate and validate a function for the output delay, to
give solid evidence for the appropriateness of using a \gls{st} for
metastability filtering in the value domain, and to elaborate on the benefits of
limiting the dynamic range of $V_{in}$.

Limitations lie in idealizations made in the process of modeling, like the
first-order approach for the dynamic behavior, ignoring parasitics, noise and
the curved shape of the $\gamma_2$ line. In our simulations we have found
confirmation that the errors thus introduced are acceptable and therefore the
key effects are well reflected in the model, but more details should be explored
here, especially for new technologies.


\end{document}